\documentclass[pdflatex,sn-aps]{sn-jnl}

\usepackage{graphicx}%
\usepackage{multirow}%
\usepackage{amsmath,amssymb,amsfonts}%
\usepackage{amsthm}%
\usepackage{mathrsfs}%
\usepackage[title]{appendix}%
\usepackage{xcolor}%
\usepackage{textcomp}%
\usepackage{manyfoot}%
\usepackage{booktabs}%
\usepackage{algorithm}%
\usepackage{algorithmicx}%
\usepackage{algpseudocode}%
\usepackage{listings}%

\theoremstyle{thmstyleone}%

\theoremstyle{thmstyletwo}%

\theoremstyle{thmstylethree}%

\begin{document}

\title[On the properties of qudits]{On the properties of qudits}

\author*[1]{\fnm{A. B.} \sur{Balantekin}}\email{baha@physics.wisc.edu}
\author[1,2,3]{\fnm{Anna M.} \sur{Suliga}}\email{asuliga@berkeley.edu}

\affil*[1]{\orgdiv{Department of Physics}, \orgname{University of Wisconsin--Madison}, \orgaddress{\city{Madison}, \postcode{53706}, \state{Wisconsin}, \country{USA}}}

\affil[2]{\orgdiv{Department of Physics}, \orgname{University of California Berkeley}, \orgaddress{\city{Berkeley}, \postcode{94720}, \state{California}, \country{USA}}}

\affil[3]{\orgdiv{Department of Physics}, \orgname{University of California, San Diego}, \orgaddress{\city{La Jolla}, \postcode{92093-0319}, \state{California}, \country{USA}}}

\abstract{Motivated by the growing interest in the applications of quantum information science in astrophysical settings, especially for the neutrino transport in compact objects where three-flavors of neutrinos need to be mapped on qutrits, we review properties of one- and two-qudit systems. We contrast two-qubit and two-qudits systems by pointing out how some of the properties of two-qubit systems generalize to higher dimensions and explore emerging new properties for dimensions three or higher. One example is provided by the Werner states: when the density operator is written in the fundamental representation, we show that only two-qubit Werner states can be pure states, but not two-qudit Werner states when the qudit dimension is larger than two.} 
                             
\maketitle

\section{Introduction}
\label{Intro}

There is progressively increasing interest in using the tools of quantum information science and quantum computing in high energy~\cite{Bauer:2022hpo} and nuclear~\cite{Beck:2023xhh} physics, including the physics and astrophysics of neutrinos~\cite{Balantekin:2023qvm} (Some specific examples include Refs.~\cite{Ciavarella:2021nmj, Gustafson:2021qbt, Gonzalez-Cuadra:2022hxt,Zache:2023cfj,Calajo:2024qrc} for high energy and Ref.~\cite{Illa:2023scc} for nuclear physics). Of particular interest is simulating the neutrino transport in core-collapse supernovae and neutron-star mergers where the very large number of neutrinos present make a full many-body calculation impossible on classical computers. For those cases quantum computing is likely to be the appropriate approach. If we assume that it is sufficient to consider only two neutrino flavors, an assumption which can only be justified in a limited number of special cases, one can map those two flavors onto the two states of a qubit. Preliminary analyses of the so-called collective neutrino oscillations (where coherent forward neutrino-neutrino interactions can no longer be ignored) have indeed been carried out on the noisy intermediate-scale quantum devices utilizing qubits~\cite{Yeter-Aydeniz:2021olz, Hall:2021rbv,Illa:2022jqb, Amitrano:2022yyn, Siwach:2023wzy,Martin:2023gbo,Illa:2022zgu,Bhaskar:2023sta,Jha:2022yik}. Preliminary attempts to address collective neutrino oscillations with three flavors on quantum computers exist \cite{Siwach:2022xhx, Balantekin:2023qvm,Nguyen:2022snr}.  
One purpose of this paper is to lay out the formal framework for the systematic study of three-flavor case. 
Since there are already many references to multiqudit systems in the literature \cite{2016JPhA...49p5203G,2016PhRvA..93f2126K,kummer2001theory,2016PhRvA..93f2320G,2008JPhA...41w5303B,Ciavarella:2021nmj, Gustafson:2021qbt, Gonzalez-Cuadra:2022hxt,Zache:2023cfj,Calajo:2024qrc,Illa:2023scc,2020FrP8479W} the current paper should be viewed as complementary to those studies. We also refer to several excellent textbooks for more detailed information on Lie algebras and Lie groups~\cite{Pfeifer,Georgi}.

An immediate question is choosing an appropriate parametrization for the density matrix \cite{2006JPhA...39.5921K}. Three possible bases are 
the generalized Gell–Mann matrix basis, the polarization operator basis and the Weyl operator basis. In this paper we choose the Gell-Mann matrix basis, the procedures for converting into other bases are available in the literature \cite{2008JPhA...41w5303B}. 

This paper is organized as follows. In the Sec.~\ref{sec:multilevel} we establish our notation by introducing properties of density matrices, entanglement entropy, and connection to the elementary symmetric polynomials (Sec.~\ref{sec:properties_qudits}). In the same section we work out qutrit properties as a non-trivial example of qudits beyond qubits (Sec.~\ref{sec:Ex_single_qutrit}). The Sec.~\ref{sec:two-qudits} is devoted to the two-qudit systems. We first summarize several results scattered through the literature for two-qubit systems and then list some results previously not commented on in the literature (Sec.~\ref{sec:two_qubits}). Subsequently, we explore two-qudit systems, point out how the properties of two-qubit systems generalize to higher dimensions and also comment on the emerging new properties for dimensions three or higher (Sec.~\ref{sec:two_qudits}). We conclude and discuss the main points presented in the work in Sec.~\ref{sec:Conclusions}. 

In the appendices we list the properties of the symmetric polynomials of the characteristic equation in App.~\ref{ApA},
we outline the connection between fundamental and adjoint representations of SU(N) in App.~\ref{apB}, and we show the relations between Gell-Mann matrices and direct products of Pauli matrices for SU(4) in App.~\ref{ApC}.

\section{Multilevel systems and qudits}
\label{sec:multilevel}

In this section, we first describe the formalism used to treat multilevel systems and expand on some of the qudits properties in Sec.~\ref{sec:properties_qudits}. Following, in Sec.~\ref{sec:Ex_single_qutrit} we discus the implications of those properties on the single qutrit example. 

\subsection{Properties of qudits}
\label{sec:properties_qudits}

The density matrix for a given $N$ level system or a qudit can be written in various bases. In the basis of generalized Gell-Mann matrices $\lambda_a$ we can write this density operator as
\begin{equation}
\label{rho} 
\rho = \frac{1}{N} ( \mathbf{1} + \lambda_a P_a) \ ,
\end{equation}
where $\mathbf{1}$ is the $N \times N$ unit matrix and $P_a$ is a component of a real vector with $N^2-1$ entries sitting in the adjoint representation of the SU(N) group. 
The generalized Gell-Mann matrices are Hermitian and they satisfy the equation 
\begin{equation}
\label{lambdef}
\lambda_a \lambda_b = \frac{2}{N} \delta_{ab} \mathbf{1} + \left( d_{abc} +i f_{abc} \right) \lambda_c \ .
\end{equation}
In the above equation $f_{abc}$ are the structure constants of the SU(N) algebra,
\begin{equation}
\label{alge}
[\lambda_a, \lambda_b] = 2 i f_{abc} \lambda_c \ , 
\end{equation}
and $d_{abc}$ is a basis-dependent tensor which is completely symmetric under exchange of its indices. It is given as
\begin{equation}
\label{ddef}
d_{abc} = \frac{1}{4} {\rm Tr} \> (\{\lambda_a, \lambda_b\} \lambda_c) \ .
\end{equation}
All $d_{abc}$ vanish for SU(2).

\subsubsection{Properties of the Density Matrix}
\label{sec:density_matrix_properties}
A density matrix should satisfy four conditions:
\begin{enumerate}[i.]
    \item{It is Hermitian,}
    \item{Its trace is one,}
    \item{It is positive semi-definite (i.e., all its eigenvalues should be positive or zero),}
    \item{$\rho^2 \le \rho$.}
\end{enumerate}
Mathematically the fourth condition is not an independent one, but it often provides a useful check of consistency. 
The density matrix in Eq.~\eqref{rho} satisfies the first two conditions by construction. The third condition implies that only certain values of $P_a$ are permissible. One way to ensure that the density matrix is positive semidefinite is to show that elementary symmetric polynomials formed from its eigenvalues are non-negative (for a proof, see, e.g., Ref.~\cite{2003PhLA..314..339K}). 
There are only $N$ such non-zero polynomials for an $N \times N$ matrix and they can be calculated by evaluating the traces of the powers of $\rho$. We list the properties of these polynomials in App.~\ref{ApA}. 

The density matrix in Eq.~\eqref{rho} can represent either a pure state or a mixed state. For a pure state it should satisfy the additional condition $\rho^2 = \rho$ and for a mixed state $\rho^2 < \rho$. 
Imposing the pure-state condition $ \rho^2 = \rho$ one gets 
\begin{equation}
\label{peq}
|P|^2 = \frac{N(N-1)}{2} \ ,
\end{equation}
and
\begin{equation}
\label{Pqequ}
\left( 1 - \frac{2}{N} \right) P_a = \frac{1}{N} d_{bca} P_bP_c \ .
\end{equation}
Hence the norm squared of the vector $P$ could lie on a $N^2 -1$-dimensional hypersphere~\footnote{Some authors in the literature rescale $P$ so that $|P|^2$ is always one. We keep the $N(N-1)/2$ factor to emphasize its scaling with $N$.}, but not all points on that hypersphere correspond to a pure state because of the Eq.~\eqref{Pqequ}. Note that multiplying Eq.~\eqref{Pqequ} with $N P_a$ and summing over $a$ we then get 
\begin{equation}
\label{qeq}
Q = d_{bca} P_b P_c P_a  = \frac{N(N-1)(N-2)}{2} \ ,
\end{equation}
for a pure state. 

Clearly for a pure state the condition $\rho = \rho^k$, where $k$ is an arbitrary integer, implies that $\rho$ is a positive semidefinite matrix. Indeed inserting this condition into Eqs.~\eqref{elsymm1} through \eqref{elsymm6} and using ${\rm Tr} \>\rho = 1$ we see that all the elementary symmetric polynomials of the eigenvalues of such a density matrix are non-negative. Even when a state is not pure, its  positive semidefiniteness implies that ${\rm Tr} \> \rho^k \ge 0$ for all $k$. Positivity of the elementary symmetric polynomials of the eigenvalues of the density matrix has a physical interpretation. The positivity of $e_2$ of Eq.~\eqref{elsymm2} implies 
\begin{equation}
\label{Eq:8}
e_2 \ge 0 \Rightarrow  {\rm Tr} \rho^2 \le 1 \ .
\end{equation}
It can be shown that positivity of $e_k$ implies ${\rm Tr} \rho^k \le 1$, with equality for only pure states.  For example for $e_3$ this can be seen by setting ${\rm Tr} \rho^2 = 1 - \epsilon$ with $0 \le \epsilon \le 1$ and 
${\rm Tr} \rho^3 = 1 - \delta$. Eq.~\eqref{elsymm3} then yields 
\begin{equation}
\label{eq:epsilon_condition}
\epsilon \ge \frac{2}{3} \delta \ ,   
\end{equation}
implying that ${\rm Tr} \rho^3 \le 1$.

\subsubsection{Entanglement Entropy}
\label{sec:entangelment_entropy}

The entanglement entropy is given by 
\begin{equation}
\label{entropy}
S = - {\rm Tr} \> \rho \log \rho \ .
\end{equation}
It vanishes for a pure state.  
The characteristic equation for a $k \times k$ density matrix is 
\begin{equation}
e_k - e_{k-1} \rho + e_{k-2} \rho^2 - \cdots (+ \> {\rm or} \> -) \rho^k = 0 \ , 
\end{equation}
implying that the eigenvalues of the density matrix and hence the entanglement entropy can be written as  functions of the elementary symmetric polynomials of the eigenvalues of the density matrix. We calculate the first few to illustrate the duality to the Casimir operators of the SU(N). We have 
\begin{eqnarray}
e_ 2 & =  &\frac{N-1}{2N} - \frac{1}{N^2} |P|^2 \ , \\
 e_3 & = &\frac{(N-1)(N-2)}{6N^2}-\frac{N-2}{N^3} |P|^2 +  \frac{2}{3N^3} Q \ , \\
e_4 & =&  \frac{(N-1)(N-2)(N-3)}{24 N^3} - \frac{(N-2)(N-3)}{2N^4} |P|^2 + \frac{2(N-3)}{3N^4}Q \\ &+& \frac{1}{2N^4} |P|^4  - \frac{1}{2N^4} \left( \frac{2}{N} |P|^4 + q_a q_a \right)  
\end{eqnarray}
In the equations above we defined 
\begin{equation}
q_a = d_{abc}P_b P_c \ ,
\end{equation}
and $Q$ is given by Eq.~\eqref{qeq}.
Hence the entanglement entropy depends on the quantities $P_a P_a, Q = d_{abc} P_a P_b P_c, d_{abc}d_{aef} P_b P_c P_e P_f, \cdots$, which are dual to the SU(N) Casimir operators $F_a F_a, d_{abc} F_a F_b F_c, d_{abc}d_{aef} F_b F_c F_e F_f, \cdots$ where $F_a$ are elements of the SU(N) algebra.

We next consider transformations of the  density matrix in Eq.~\eqref{rho} under the SU(N) transformations:
\begin{equation}
\label{transform}
\rho \rightarrow \rho' = U \rho U^{\dagger}. 
\end{equation}
Since the density matrix is written in terms of the fundamental representation of the SU(N) algebra, $U$ in the above equation is in the fundamental representation of the SU(N) group. Note that this transformation leaves the entanglement entropy invariant. Using Eq.~\eqref{untersi} of App.~\ref{apB} in Eq.~\eqref{transform} we get 
\begin{equation}
 U \rho U^{\dagger} = \frac{1}{N}(\mathbf{1}+ R_{ki}P_i \lambda_k) \ ,
\end{equation}
or
\begin{equation}
P_i \rightarrow P_i'=R_{ij} P_j \ .
\end{equation}
In these equations $R_{ij}$ are the matrix elements of the adjoint representation of SU(N) (see App.~\ref{apB}). 
The Eq. (\ref{orthogonality}) implies 
\begin{equation}
P_i P_i = P_j' P_j' \ .
\end{equation}
Similarly Eq.~\eqref{dequation} implies 
\begin{equation}
d_{ijk} P_i P_j P_k = d_{abc}P_a' P_b' P_c' \ .
\end{equation}
Continuing in this fashion one can show that all the quantities entanglement entropy depends on remain unchanged under SU(N) transformations. Here we provided the proof when the density matrix is written in the fundamental representation of SU(N), the subject of this review. It is straightforward to generalize this proof to any representation.

\subsection{Example: single qutrit}
\label{sec:Ex_single_qutrit}

For a single qutrit we first consider the matrix ${\cal A} = \lambda_a P_a$ where $\lambda_a$ are the SU(3) Gell-Mann  matrices: 
\begin{equation}
\rho = \frac{1}{3} (\mathbf{1} + {\cal A}) \ .
\end{equation}
Although the positive definiteness condition is automatically satisfied for the single-qubit density matrices, that is no longer the case for qudits of dimension $N$ for $N\ge 3$~\cite{Deen:1971hfy}. As mentioned above, the elementary symmetric polynomials of the eigenvalues of the traces of first three powers of these density matrices should be non-negative, see discussion in Sec.~\ref{sec:density_matrix_properties}. 
The conditions that the SU(3) invariants satisfy for a single qutrit are
\begin{subequations}
    \begin{equation}
 |P|^2 \le 3 \ ,
    \end{equation}
\begin{equation}
   \frac{2}{3} Q \ge |P|^2 - 1 \ .
    \label{condition1}
\end{equation} 
\end{subequations}
However these inequalities do not completely specify the necessary restrictions since all positive values can be realized. It is also necessary to examine the roots of the characteristic equation~\cite{Deen:1971hfy}. 

It is faster to examine the characteristic equation for the matrix ${\cal A}$ only. We have 
\begin{equation}
{\rm Tr} \> {\cal A} = 0 \ , \nonumber 
\end{equation}
\begin{equation}
{\rm Tr} \> {\cal A}^2 = 2 |P|^2 \ , \nonumber
\end{equation}
and
\begin{equation}
{\rm Tr} \> {\cal A}^3 = 2Q \ . \nonumber
\end{equation}
Hence the characteristic equation for the matrix ${\cal A}$ is 
\begin{equation}
\label{eq:char_pol_3}
x^3 -  |P|^2 x - \frac{2}{3} Q = 0 \ , 
\end{equation}
i.e., the entanglement entropy depends only on $|P|$ and $Q$. 
As mentioned before, the density matrix should be positive semidefinite. This requires all roots of this equation to be real and take values so that $(1+x_i)$ are positive. If all the $x_i$ are real then the discriminant  Eq.~\eqref{eq:char_pol_3} is positive:
\begin{equation}
-4 (-|P|^6) - 27 \frac{4}{9} Q^2 > 0 \ , 
\end{equation}
or
\begin{equation}
3 \frac{Q^2}{|P|^6} <1 \ . 
\end{equation}
This suggests that 
\begin{equation}
\label{condition2}
-1 \le \sqrt{3} \frac{Q}{|P|^3} \le +1 \ ,
\end{equation}
so that it can be written as trigonometric functions. The roots of the characteristic equation are given by
\begin{subequations}
\label{eq:eigenvalues} 
\begin{eqnarray}
x_1 &=& \frac{2 |\vec{P}|}{\sqrt{3}} \left[ -\frac{1}{2}\cos \left(\frac{\chi}{3}\right) -\frac{\sqrt3}{2} \sin \left(\frac{\chi}{3}\right) \right] \ , 
\label{eq:x1} \\
x_2 &=& \frac{2 |\vec{P}|}{\sqrt{3}} \left[ -\frac{1}{2}\cos \left(\frac{\chi}{3}\right) +\frac{\sqrt3}{2} \sin \left(\frac{\chi}{3}\right) \right] \ ,
\label{eq:x_2} \\
x_3 &=& \frac{2 |\vec{P}|}{\sqrt{3}} \cos\left(\frac{\chi}{3}\right) \ ,  
\label{eq:x3}
\end{eqnarray}
\end{subequations}
where
\begin{equation}
\cos \chi = \sqrt{3} \frac{Q}{|P|^3} \ .
\end{equation}
%

\begin{figure}[t]
\centering
 	\includegraphics[width=0.6\columnwidth]{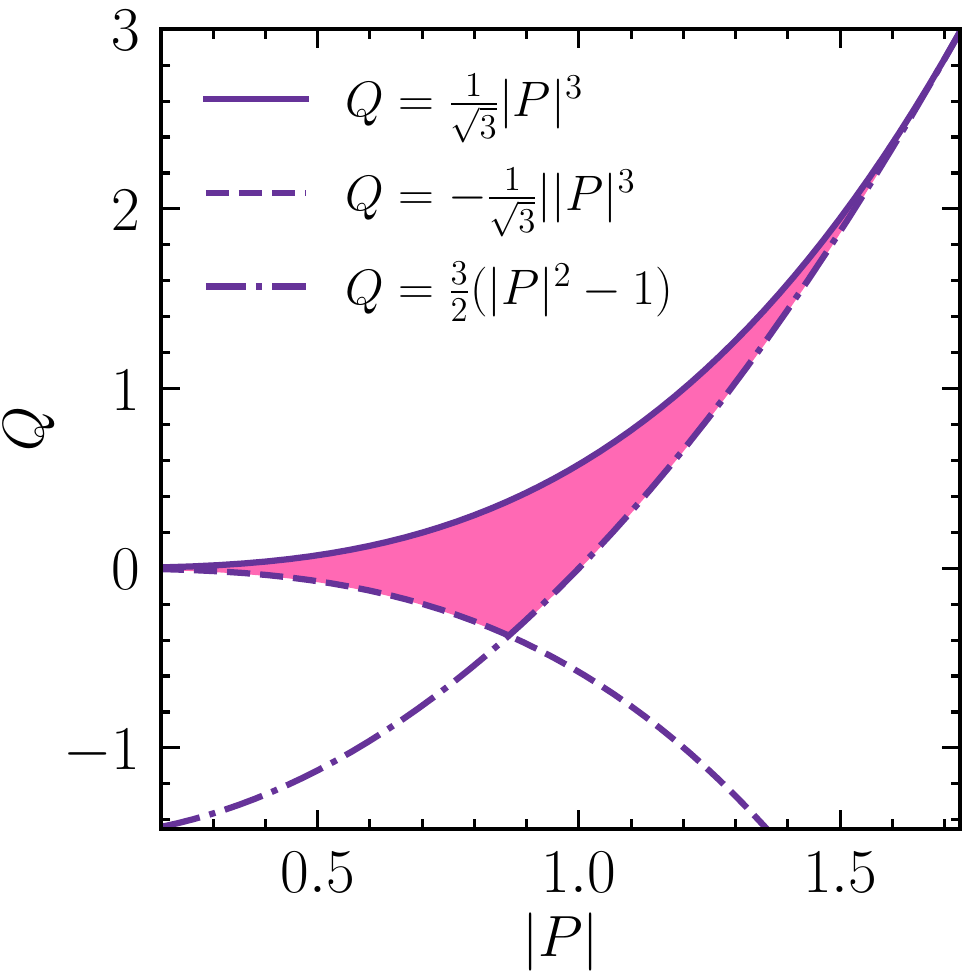}
	\caption{Shaded area: allowed values of $Q$ and $|P|$ for a qutrit imposing the positivity conditions. The dot-dashed line represents the condition given in Eq. (\ref{condition1}) and solid and dashed lines represent the conditions given in Eq.~(\ref{condition2}). Note that the upper right corner of the figure corresponds to a pure state.}
	\label{Fig:Qubit_regions}
\end{figure}

Clearly not all values in the range given in Eq.~\eqref{condition2} yield positive definite eigenvalues of the density matrix (see also Ref.~\cite{2000OptCo.179..439C}). The conditions given in Eq.~\eqref{condition1} also need to be satisfied. The allowed values of $Q$ and $|P|$ are displayed in Fig.~\ref{Fig:Qubit_regions}. It is apparent from the graph that, unlike $|P|^2$, $Q$ can take negative values.

\hskip 20pt
\section{Two-qudit systems} 
\label{sec:two-qudits}

In this section, we examine the properties of two-qubit (Sec.~\ref{sec:two_qubits}) and two-qudit systems (Sec.~\ref{sec:two_qudits}). We specifically comment on how the properties inherent to two-qubit systems extend to higher dimensions and which properties emerge only in dimension three or higher.

\subsection{Two-qubits}
\label{sec:two_qubits}

Study of interacting a couple of two-level systems have a long history, for an earlier review see, e.g., Ref~\cite{Fano:1983zz}. 
In general density matrices for one qudit with $N=4$ or a system of two qubits can be written using SU(4) generators. 
In the previous section we used the Gell-Mann (or generalized Gell-Mann) representation of these generators. (For a description of different bases for SU(N) and the procedure to 
write down Generalized Gell-Mann matrices see e.g. Ref.
\cite{2008JPhA...41w5303B}). 
Especially for two qubit systems, either non-interacting (pure) or interacting (mixed), it is physically more transparent to write the SU(4) generators in terms of Pauli matrices. (The relation between Gell-Mann matrices and direct products of Pauli matrices forming the SU(4) basis is given in Appendix \ref{ApC}).
The two-qubit density matrix then takes the form 
\begin{equation}
\label{twoneutrino}
\rho_{2q} = \frac{1}{4} \left[ (\mathbf{1} \otimes \mathbf{1}) + (\sigma_i \otimes \mathbf{1}) x_i + (\mathbf{1} \otimes \sigma_i) y_i + (\sigma_i \otimes \sigma_j) \omega_{ij} \right]
\end{equation}
Here $x_i$ and $y_i$ are three-vectors and $\omega_{ij}$ is a $3 \times 3$ real matrix. If this density matrix is not describing a pure state $x_i, y_i, {\rm and} \> \omega_{ij}$ need to satisfy further conditions we give at the end of this subsection. 

We next quantify the condition for this density matrix to represent a pure state.  We first calculate 
\begin{multline}
    \rho_{2q}^2 = \frac{1}{16} \left\{ (\mathbf{1} \otimes \mathbf{1}) (1 + |\mathbf{x}|^2 + |\mathbf{y}|^2 + \omega_{ij} \omega_{ij} )   \right. + (\sigma_i \otimes \mathbf{1}) (2 x_i + 2 \omega_{ij} y_j) 
+ (\mathbf{1} \otimes \sigma_i) (2 y_i + 2 x_j \omega_{ji} )  \\
+ \left. (\sigma_i \otimes \sigma_j) \left[ 2 \omega_{ij} + x_i y_j - \delta_{ij} \left( ({\rm Tr}\> \omega)^2 - {\rm Tr} \> \omega^2 \right) 
+ 2 \omega_{ji} {\rm Tr} \> \omega - 2 (\omega^2)_{ji} \right] \right\} \ .
\end{multline}
To satisfy the pure state condition $\rho^2 = \rho$ we need 
\begin{equation}
\label{bir}
 1 + |\mathbf{x}|^2 + |\mathbf{y}|^2 + \omega_{ij} \omega_{ij} = 4 \ , 
\end{equation} 
\begin{equation}
\label{iki}
x_i = \omega_{ij} y_j \ ,
\end{equation}
\begin{equation}
\label{uc} 
y_i = x_j \omega_{ji} \ ,
\end{equation}
\begin{equation}
\label{dort}
\omega_{ij} = x_iy_j - \frac{1}{2} \delta_{ij} \left[ ({\rm Tr} \> \omega)^2 - {\rm Tr} \> \omega^2 \right] + \omega_{ji} ({\rm Tr} \> \omega) - (\omega^2)_{ji} \ . 
\end{equation} 
Taking the trace of Eq.~\eqref{dort} one obtains 
\begin{equation}
\label{bess}
{\rm Tr}\> \omega = \mathbf{x} \cdot \mathbf{y} - \frac{1}{2} \left[ 
({\rm Tr}\> \omega)^2 - {\rm Tr}\> \omega^2 \right] \ , 
\end{equation}
i.e., Eq.~\eqref{bess} provides the relationship between first two elementary symmetric functions of eigenvalues of the matrix $\omega$.  
Multiplying Eq.~\eqref{iki} with $x_i$, summing over $i$ and using Eq.~\eqref{uc} we see that
\begin{equation}
\label{equal}
x_ix_i = x_i \omega_{ij} y_j = y_jy_j \>\>\> \Rightarrow |\mathbf{x}|^2 = |\mathbf{y}|^2 \ . 
\end{equation}
Multiplying Eq.~\eqref{dort} with $\omega^T$ we get
\begin{multline}
(\omega \omega^T)_{ik} = x_ix_k - \frac{\omega_{ki}}{2} \left[ ({\rm Tr} \> \omega)^2 - {\rm Tr} \> \omega^2 \right]  + (\omega^2)_{ki} ({\rm Tr} \> \omega) - (\omega^3)_{ki} \ ,
\end{multline}
where we used Eq.~\eqref{iki} in rewriting the first term in the right side of the equation. Taking the trace of both sides we obtain
\begin{multline}
{\rm Tr}\>(\omega \omega^T)  = |\mathbf{x}|^2 -\frac{1}{2} ({\rm Tr} \> \omega)  \left[ ({\rm Tr} \> \omega)^2 - {\rm Tr} \> \omega^2 \right]  + ({\rm Tr} \> \omega^2) ({\rm Tr} \> \omega) 
- {\rm Tr} \> \omega^3 \ .
\end{multline}
Using the relations between power sums and elementary symmetric polynomials given in the Eqs.~\eqref{eq:A1} - \eqref{eq:A2} in App.~\ref{ApA} 
and that ${e_3} = \det \omega$ we see that
\begin{equation}
\label{deter}
{\rm Tr}\>(\omega \omega^T)  = |\mathbf{x}|^2 - 3 e_3 =  |\mathbf{x}|^2 - 3 \det \omega \ .
\end{equation}
Substituting Eqs.~\eqref{equal} and (\ref{deter}) into Eq.~\eqref{bir} we get that 
\begin{equation}
\label{detconnection}
 |\mathbf{x}|^2 = |\mathbf{y}|^2 = 1 + \det \omega \ , 
\end{equation}
i.e., for a physical state $\det \omega \le 0$ (zero for a pure state). Next consider the Cayley-Hamilton theorem which states that  the matrix $\omega$ satisfies its characteristic polynomial:
\begin{equation}
\label{eq:char_eq}
\omega^3 -\omega^2 e_1 + \omega e_2 - \mathbf{1}e_3 = 0  \ , 
\end{equation}
where 
$e_1 = \rm{Tr}(\omega)$ and $e_2 = 1/2(\rm{Tr}\left(\omega)\right)^2- 1/2\rm{Tr}(\omega^2)$.

Defining the matrix 
\begin{equation}
\label{char}
Z_{ij} = - \frac{1}{2} \delta_{ij} \left[ ({\rm Tr} \> \omega)^2 - {\rm Tr} \> \omega^2 \right] + \omega_{ji} ({\rm Tr} \> \omega) - (\omega^2)_{ji} \ ,
\end{equation}
we see that 
\begin{equation}
\label{eq:Z_T}
Z^T = -\mathbf{1} e_2 +\omega e_1 - \omega^2 \ .
\end{equation}
Multiplying this equation with $\omega$ we obtain
\begin{equation}
\omega Z^T = - \omega e_2 + \omega^2 e_1 - \omega^3 \ .
\end{equation}
Comparing this with Eq.~\eqref{eq:char_eq} we finally obtain
\begin{equation}
\label{Zmatrix}
\omega Z^T = - \det{\omega}\> \mathbf{1} \Rightarrow \> Z^T = -  \det{\omega} \> \omega^{-1}. 
\end{equation}
In other words $-Z^T$ is the adjugate matrix of $\omega$. Note that Eq. (\ref{dort}) can be written as 
\begin{equation}
\omega_{ij} = x_i y_j + Z_{ij} .
\end{equation}
If Eq.~\eqref{twoneutrino} were to describe two unentangled qubits $\omega_{ij}$ would be $x_i y_j$. Hence a non-zero $Z$ quantifies the entanglement between the two qubits we are considering. $Z$ is a null matrix only when $ \det{\omega}=0$. Indeed Eq.~\eqref{detconnection} shows that in that case the two qubits are unentangled ($|\mathbf{x}|^2 = |\mathbf{y}|^2 = 1$). 

If the density matrix in Eq.~\eqref{twoneutrino} represents a pure state the conditions given in Eqs.~\ref{bir} - \eqref{dort} are sufficient for it to be positive semi-definite. But if it represents a state which is not pure, then the elementary symmetric polynomials of the eigenvalues of the traces of first four powers of this density matrix should be less than one and non-negative. The first one, ${\rm Tr} \> \rho$, is positive by construction. The conditions other three need to satisfy can be calculated using the expressions in App.~\ref{ApA}. We get 
\begin{subequations}
\begin{equation}
 |{\mathbf x}|^2+|{\mathbf y}|^2 +\omega_{ij} \omega_{ij} \le 3 ,
\end{equation}
\begin{equation}
|{\mathbf x}|^2+|{\mathbf y}|^2 +\omega_{ij} \omega_{ij} \le \frac{1}{9} +
2(x_i \omega_{ij}x_j - \det \omega ) ,
\end{equation}
\begin{multline}
1 - 2 (|{\mathbf x}|^2+|{\mathbf y}|^2 +\omega_{ij} \omega_{ij}) + 
(|{\mathbf x}|^2+|{\mathbf y}|^2 +\omega_{ij} \omega_{ij})^2 \\ + 8 (x_i \omega_{ij}x_j - \det \omega) - 4x_i (\omega \omega^T)_{ij} x_j \\ - 4 y_i (\omega^T \omega)_{ij} y_j 
-4 ( |{\mathbf x}|^2 |{\mathbf y}|^2 + 2 y_i Z_{ij}x_j + Z_{ij} Z_{ij}) \ge 0, 
\end{multline}
\label{positivity}
\end{subequations}
where the matrix $Z$ is given by Eq.~\eqref{Zmatrix}. 
In Eqs.~\eqref{positivity} the equality sign holds only when the state in Eq.~\eqref{twoneutrino} is a pure state.

\subsection{Two qudits} 
\label{sec:two_qudits}

For a system of two dimension-$N$ qudits one can write the density matrix in terms of the generators of the SU(N$^2$) algebra expressed as direct products of the generators of the SU(N) algebra: 
\begin{equation}
\label{twoqudit}
    \rho_{2q} = \frac{1}{N^2} [ (\mathbf{1} \otimes \mathbf{1}) +  (\lambda_i \otimes \mathbf{1}) x_i +   (\mathbf{1} \otimes \lambda_i) y_i +  (\lambda_i \otimes \lambda_j) \omega_{ij} ] \ .
\end{equation}
If we want the density matrix in Eq.~\eqref{twoqudit} to represent a pure state, imposing $\rho = \rho^2$, we get four conditions:
\begin{subequations}
\begin{equation}
\label{eq:sum}
    N^2 = 1 + \frac{2}{N}(|\mathbf{x}|^2 + |\mathbf{y}|^2) + \frac{4}{N^2} \omega_{ij}\omega_{ij} \ ,
\end{equation}
\begin{equation}
\label{eq:xi}
    (N^2-2)x_i = d_{kji} x_k x_j + \frac{4}{N}y_j\omega_{ij} + \frac{2}{N}d_{mki}\omega_{ml}\omega_{kl} \ ,
\end{equation}
\begin{equation}
\label{eq:yi}
    (N^2-2)y_i = d_{kji}y_k y_j + \frac{4}{N}x_j\omega_{ji} + \frac{2}{N}d_{mki}\omega_{lm}\omega_{lk} \ ,
\end{equation}
\begin{equation}
\label{quditomeg}
   (N^2-2) \omega_{ij} = 2 x_i y_j  + 2d_{kli} x_k \omega_{lj}   + 2 d_{klj}y_k \omega_{il}  + C_{ij} \ ,
\end{equation}
where 
\begin{equation}
    C_{ij} = \omega_{mn}\omega_{kl}\left(  d_{nlj}d_{mki} - f_{nlj}f_{mki}  \right) \ .
\end{equation}
\end{subequations}

Defining $z_i = d_{imk}\omega_{mk}$ and using the SU(N) identity \cite{Haber:2019sgz}
\begin{equation}
f_{mki} f_{nli} = \frac{2}{N} \left( \delta_{mn} \delta_{kl} - \delta_{ml} \delta_{kn} \right) + d_{mni} d_{kli} - d_{kni} d_{mli} \ ,
\end{equation}
Eq.~\eqref{quditomeg} yields 
\begin{eqnarray} 
(N^2 -2) {\rm Tr} \> \omega &=&    2 \mathbf{x} \cdot \mathbf{y} + 2 (\mathbf{x} + \mathbf{y}) \cdot \mathbf{z} \nonumber \\ &+& \frac{1}{2} d_{imk} d_{inl} (\omega + \omega^T)_{mn}  (\omega + \omega^T)_{kl} \nonumber \\ &-& \frac{2}{N} \left( ({\rm Tr} \> \omega)^2- {\rm Tr} \> \omega^2 \right) - {\mathbf z}^2 \ . 
\label{omegtrace}
\end{eqnarray}
Eq.~\eqref{omegtrace} generalizes Eq.~\eqref{bess} from qubits to qudits. 

Taking partial traces of the density matrix in Eq.~\eqref{twoqudit} we obtain the reduced density matrices for each qubit:
\begin{subequations}
\begin{equation}
\rho_1 = {\rm Tr_2} \rho = \frac{1}{N}(\mathbf{1} + \lambda_i x_i) \ ,
\end{equation}
\begin{equation}
\rho_2 = {\rm Tr_1} \rho = \frac{1}{N}(\mathbf{1} + \lambda_i y_i) \ .
\end{equation}
\end{subequations}
These reduced density matrices should each be also positive semidefinite. This brings additional conditions. 
(For qutrits we explicitly discussed these conditions in Sec.~\ref{sec:Ex_single_qutrit}). 

Note that the Eqs.~\eqref{eq:sum} through \eqref{quditomeg} and Eq. ~\eqref{omegtrace} are invariant under the simultaneous transformations $x_i \leftrightarrow y_i$ and $\omega \leftrightarrow \omega^T$. Using Eqs.~\eqref{peq} and \eqref{Pqequ} it is also straightforward to show that tensor product of two pure states, i.e, $\omega_{ij} = x_i y_j$, satisfies these equations. 

A Werner state corresponds to a density matrix representing a bipartite system which remains invariant under the transformation by all the unitary operators in the form $U \otimes U$ \cite{Werner:1989zz}: 
\begin{equation}
\label{wernerdef}
\rho = (U \otimes U) \rho (U^{\dagger} \otimes U^{\dagger}) \ .
\end{equation}
For two dimension-$N$ qudit density matrix in Eq.~\eqref{twoqudit} to be Werner state density matrix $U$ in Eq.~\eqref{wernerdef} should be the most general SU(N) transformation. We get 
\begin{eqnarray}
\label{twoqudittr}
\rho_{2q} &=&  (U \otimes U) \rho_{2q} (U^{\dagger} \otimes U^{\dagger}) \nonumber  \\ &=& \frac{1}{N^2} \left[ (\mathbf{1} \otimes \mathbf{1}) +  (U\lambda_i U^{\dagger} \otimes \mathbf{1}) x_i +   (\mathbf{1} \otimes U \lambda_i U^{\dagger}) y_i   +  (U \lambda_i U^{\dagger}\otimes U \lambda_j U^{\dagger}) \omega_{ij} \right] \ . 
\end{eqnarray}
Using Eq.~\eqref{untersi} of App.~\ref{apB} one can get the conditions
\begin{subequations}
\begin{equation}
R_{ij} x_i = x_j, \>\>\>\> R_{ij} y_i = y_j \ ,
\end{equation}
\begin{equation}
\omega_{nm} = R^T_{ni} \omega_{ij} R_{jm} \ ,
\end{equation}
\end{subequations}
for the equality in Eq.~\eqref{twoqudittr} to be satisfied. For the conditions in these equations to be valid for any adjoint representation of SU(N) group element one needs 
\begin{equation}
\label{wernervalues}
x_i = 0 = y_i, \>\>\> \omega_{ij} = \alpha \delta_{ij} \>\> \forall \> i,j \ ,
\end{equation}
where $\alpha$ is a real number to be determined. (The condition on the $\omega$ matrix follows from the Schur's lemma). 

For a pure Werner state, using the SU(N) property $\sum_j d_{ijj}=0$ \cite{Haber:2019sgz}, it follows that Eqs.~\eqref{eq:xi} and \ref{eq:yi} are readily satisfied for the values in Eq.~\eqref{wernervalues}. Using ${\rm Tr} \> \omega \omega^T = \alpha^2 (N^2 -1)$ Eq.~\eqref{eq:sum} determines the value of $\alpha$ to be
\begin{equation}
\label{alpha1}
    \alpha^2 = \frac{N^2}{4} \ . 
\end{equation}
For a pure Werner state Eq.~\eqref{quditomeg} also need to be satisfied, i.e. the equation 
\begin{equation}
\label{lastcheck}
(N^2 -2) \alpha \delta_{ij} = \alpha^2  \delta_{mn} \delta_{kl} 
\left(  d_{nlj}d_{mki} - f_{nlj}f_{mki}  \right) \ ,
\end{equation} 
also needs to hold. Using the SU(N) relations \cite{Haber:2019sgz} 
\begin{equation}
f_{ijk} f_{ijn} = N \delta_{kn} \ ,
\end{equation}
and 
\begin{equation}
d_{ijk} d_{ijn} = \left( \frac{N^2 -4}{N} \right) \delta_{kn} \ ,
\end{equation}
Eq.~\eqref{lastcheck} determines $\alpha$ to be 
\begin{equation}
\label{alpha2}
    \alpha = - \frac{N(N^2-2)}{4} \ .
\end{equation}
Eqs.~\eqref{alpha1} and \eqref{alpha2} are consistent only for $N=2$. Indeed for two qubits the Werner state can be written either as 
\begin{equation}
\rho = \frac{1}{4} \left[(\mathbf{1} \otimes \mathbf{1}) - \alpha \sum_{i=1}^3 (\sigma_i \otimes \sigma_i) \right] \ ,
\end{equation}
or as 
\begin{equation}
\rho = - \alpha | \Psi^- \rangle \langle \Psi^- | + \frac{1+\alpha}{4} (\mathbf{1} \otimes \mathbf{1}) \ ,
\end{equation}
where $| \Psi^- \rangle = ( |01 \rangle - |10 \rangle) /\sqrt{2}$ is a Bell state. Clearly a pure Werner state is only possible for $\alpha= -1$ as given by Eq.~\eqref{alpha2} for $N=2$. 

For mixed two-qudit Werner states the  positive semi-definiteness condition of the density matrix requires the elementary symmetric functions given in App.~\ref{ApA} to be all positive. For example the condition $e_2 \ge 0$, equivalent to 
the condition ${\rm Tr} \> \rho^2 < 1$ (cf. Eq. \ref{Eq:8}),  we obtain the possible ranges of $\alpha$ to be bounded as 
\begin{equation}
\label{Eq:64}
- \frac{N}{2} < \alpha < + \frac{N}{2}  \> . 
\end{equation}
It is important to emphasize that this is a {\it necessary}, but not a {\it sufficient} condition. Similarly the requirement $e_3 \ge 0$ gives the condition
\begin{equation}
    \left[(N^2-2) - 12 (N^2 -2) \left(\frac{\alpha}{N} \right)^2- 32 \left(\frac{\alpha}{N} \right)^3\right] \ge 0 ,
\end{equation}
which is again necessary, but not sufficient. For two dimension-N qudits the elementary symmetric functions $e_2, e_3, \cdots, e_{N^2}$ should all be non-negative. See also Refs.~\cite{2015PhRvA..92c2107Q, 2020PhRvA.101b2112J,2006RpMP...58..325C,2022Photo...9..741W} for more discussion of this topic.

\section{Conclusions}
\label{sec:Conclusions}

The growing field of quantum information science opens up opportunities to study many-body problems in astrophysics. Of particular interest are the neutrino-neutrino interactions in compact objects as that is the only place, except the Early Universe, where neutrino densities are large enough to permit efficiently such interactions.

In this work, we summarized our findings on properties of two-qubit systems and extend the treatment to two-qudit systems. In generalizing the two-qubit systems to higher dimensions, we point out the emergence of new properties for dimensions three or higher. 
In particular we showed that there is no pure Werner state, written in the fundamental representation of SU(N), beyond two qubits. As far as we know this was not noted in the literature before.

\bmhead{Acknowledgements}
We thank Pooja Siwach and Amol Patwardhan for discussions.
This work was supported in part by the U.S.~Department of Energy, Office of Science, Office of High Energy Physics, under Award  No.~DE-SC0019465. 
It was also supported in part by the National Science Foundation Grants  No. PHY-2020275, PHY-2108339, and Heising-Simons Foundation (Grant 2017-228). 

\phantom{i}

\appendix

\section{Symmetric Polynomials}
\label{ApA}
The characteristic equation of any matrix ${\cal M}$ with eigenvalues $x_i$ can be written as 
\begin{equation}
\label{eq:A1}
\prod_{i=1}^N (x-x_i) = \sum_{k=1}^N (-1)^k x^{N-k} e_k = 0
\end{equation}
where $e_k$ are the elementary symmetric polynomials of the eigenvalues $x_i$. 
The elementary symmetric polynomials can in turn be expressed in terms of power sums, $p_k = \sum_i x_i^k$. via the equality
\begin{equation}
\label{eq:A2}
\sum_k e_k x^k = \exp \left( \sum_k \frac{(-1)^{k+1}}{k} p_k x^k \right) \ .
\end{equation}
Note that $p_k = {\rm Tr} \> ({\cal M}^k)$. 

From the expressions above first few elementary symmetric functions of the density matrix is given as (using ${\rm Tr} \> \rho =1$)
\begin{equation}
\label{elsymm1}
e_1 = {\rm Tr} \> \rho = 1, 
\end{equation}
\begin{equation}
\label{elsymm2}
e_2 = \frac{1}{2} - \frac{1}{2} {\rm Tr} \> \rho^2, 
\end{equation}
\begin{equation}
\label{elsymm3}
e_3 =  \frac{1}{6}  - \frac{1}{2}  ({\rm Tr} \> \rho^2)+ \frac{1}{3} ({\rm Tr} \> \rho^3), 
\end{equation}
\begin{equation} 
\label{elsymm4}
e_4 = \frac{1}{24} \left( 1 - 6 \> {\rm Tr} \> \rho^2 + 3 \> ({\rm Tr} \> \rho^2)^2 + 8 \> {\rm Tr} \> \rho^3 -6 \> {\rm Tr} \> \rho^4 \right) ,
\end{equation}

\begin{equation}
\label{elsymm5}
e_5 = \frac{1}{120} \left( 1 -10 \> {\rm Tr} \rho^2 +15  ({\rm Tr}\>\rho^2)^2  + 20 \> {\rm Tr}\> \rho^3 -20 ({\rm Tr}\> \rho^2) ({\rm Tr}\> \rho^3) -30 \> {\rm Tr} \rho^4 +24 \> {\rm Tr} \rho^5 \right) 
\end{equation}
\begin{multline}
\label{elsymm6}
e_6 = \frac{1}{6!} \left( 1 - 15 \> {\rm Tr}\> \rho^2  + 45 ({\rm Tr} \> \rho^2)^2  - 15({\rm Tr} \> \rho^2)^3  40 \> {\rm Tr} \> \rho^3  - 120 ({\rm Tr} \> \rho^2) ({\rm Tr} \> \rho^3)  \right. \\ \left. + 40 \> ({\rm Tr} \> \rho^3)^2 -90 \> {\rm Tr} \> \rho^4  + 144 \> {\rm Tr} \> \rho^5   -120 \> {\rm Tr} \> \rho^6  \right)
\end{multline}

For an $N \times N$ matrix $e_k =0$ for $k>N$. Finally for the density matrix $\rho = (1 + {\cal A})/N$ we have 
\begin{equation}
{\rm Tr} \rho^k = \frac{1}{N^k} \sum_{m=0}^k \frac{k!}{m!(k-m)!} {\rm Tr} \> {\cal A}^m \ . 
\end{equation}
%

\section{Connection between fundamental and adjoint representations of SU(N)}
\label{apB}

Consider a transformation of the extended Gell-Mann matrices under the transformation of the SU(N) group, $U \lambda_j U^{\dagger}$. The result is a traceless, Hermitian matrix and as such it can be written as a linear combination of the extended Gell-Mann matrices:
\begin{equation}
 \label{untersi}
U \lambda_j U^{\dagger} = R_{kj} \lambda_k \ ,
\end{equation}
where $R_{kj}$ are real numbers. Starting with this equation it is easy to show that the matrices $R$ transform like the elements of a group. Indeed they are the elements of the $(N^2-1)$-dimensional adjoint representation of the SU(N) group. They represent the SU(N) subgroup of the SO(N$^2$-1) group. The results we outline below were derived for SU(3) in Ref. \cite{Macfarlane:1968ce}. Here we present their generalization to SU(N). 

Multiplying Eq. (\ref{untersi}) with itself we can write
\begin{equation}
U \lambda_j U^{\dagger} U \lambda_p U^{\dagger} = R_{kj} \lambda_k R_{qp} \lambda_q \ . 
\end{equation}
Rewriting the product of two lambda matrices we get 
\begin{equation}
    \frac{2}{N} \delta_{jp} + (i f_{jpz}+d_{jpz}) U \lambda_z U^{\dagger}  = \left[ \frac{2}{N} \delta_{kq} +(i f_{kqt} + d_{kqt}) \lambda_t \right] R_{kj} R_{qp} \ . 
    \label{lambprod}
\end{equation} 
Taking the trace of the above equation one obtains 
\begin{equation}
\label{orthogonality}
R^TR = 1 \ . 
\end{equation}
First removing the terms proportional to the identity matrix in Eq. (\ref{lambprod}), second using Eq. (\ref{untersi}) on the left side, third multiplying both sides with the same lambda matrix and finally taking the trace we get 
\begin{equation}
(i f_{jpz}+ d_{jpz}) R_{lz} = (i f_{kql} + d_{kql} ) R_{kj} R_{qp} \ . 
\end{equation}
Using the orthogonality of the R matrices and separating the real and imaginary parts of this equation we then obtain 
 \begin{equation}
f_{kql} R_{kj} R_{qp} R_{lm} = f_{jpm} \ , 
\end{equation}
 \begin{equation}
 \label{dequation}
d_{kql} R_{kj} R_{qp} R_{lm} = d_{jpm} \ .  
\end{equation}

\section{The relation between Gell-Mann matrices and direct products of Pauli matrices for SU(4)}
\label{ApC}

\begin{equation}
\Lambda^{12}_s = 
\left(
\begin{array}{cccc}
 0 &  1 & 0 & 0  \\
 1 &  0 & 0 & 0  \\
0  & 0  & 0 & 0 \\
0 & 0 & 0 & 0  
\end{array}
\right) = \frac{1}{2} \left( \sigma_3 \otimes \sigma_1 + \mathbf{1} \otimes \sigma_1 \right)
\end{equation}

\begin{equation}
\Lambda^{34}_s = 
\left(
\begin{array}{cccc}
 0 &  0 & 0 & 0  \\
 0 &  0 & 0 & 0  \\
0  & 0  & 0 & 1 \\
0 & 0 & 1 & 0  
\end{array}
\right) = \frac{1}{2} \left( \mathbf{1} \otimes \sigma_1 - \sigma_3 \otimes \sigma_1 \right)
\end{equation} 

\begin{equation}
\Lambda^{13}_s = 
\left(
\begin{array}{cccc}
 0 &  0 & 1 & 0  \\
 0 &  0 & 0 & 0  \\
1  & 0  & 0 & 0 \\
0 & 0 & 0 & 0  
\end{array}
\right) = \frac{1}{2} \left( \sigma_1 \otimes 1 + \sigma_1 \otimes \sigma_3 \right)
\end{equation}

\begin{equation}
\Lambda^{24}_s = 
\left(
\begin{array}{cccc}
 0 &  0 & 0 & 0  \\
 0 &  0 & 0 & 1  \\
0  & 0  & 0 & 0 \\
0 & 1 & 0 & 0  
\end{array}
\right) = \frac{1}{2} \left( \sigma_1 \otimes 1 - \sigma_1 \otimes \sigma_3 \right)
\end{equation}

\begin{equation}
\label{Lambda14s}
\Lambda^{14}_s = 
\left(
\begin{array}{cccc}
 0 &  0 & 0 & 1  \\
 0 &  0 & 0 & 0  \\
0  & 0  & 0 & 0 \\
1 & 0 & 0 & 0  
\end{array}
\right) = \frac{1}{2} \left( \sigma_1 \otimes \sigma_1 - \sigma_2 \otimes \sigma_2
\right)
\end{equation}

\begin{equation}
\Lambda^{23}_s = 
\left(
\begin{array}{cccc}
 0 &  0 & 0 & 0  \\
 0 &  0 & 1 & 0  \\
0  & 1  & 0 & 0 \\
0 & 0 & 0 & 0  
\end{array}
\right) = \frac{1}{2} \left( \sigma_1 \otimes \sigma_1 + \sigma_2 \otimes \sigma_2 \right)
\end{equation}

\begin{equation}
\Lambda^{12}_a = 
\left(
\begin{array}{cccc}
 0 &  -i & 0 & 0  \\
 i &  0 & 0 & 0  \\
0  & 0  & 0 & 0 \\
0 & 0 & 0 & 0  
\end{array}
\right) = \frac{1}{2} \left( \mathbf{1} \otimes \sigma_2 + \sigma_3 \otimes \sigma_2 \right)
\end{equation}

\begin{equation}
\Lambda^{34}_a = 
\left(
\begin{array}{cccc}
 0 &  0 & 0 & 0  \\
 0 &  0 & 0 & 0  \\
0  & 0  & 0 & -i \\
0 & 0 & i & 0  
\end{array}
\right) = \frac{1}{2} \left( \mathbf{1} \otimes \sigma_2 - \sigma_3 \otimes \sigma_2 \right)
\end{equation} 

\begin{equation}
\Lambda^{13}_a = 
\left(
\begin{array}{cccc}
 0 &  0 & -i & 0  \\
 0 &  0 & 0 & 0  \\
i  & 0  & 0 & 0 \\
0 & 0 & 0 & 0  
\end{array}
\right) = \frac{1}{2} \left( \sigma_2 \otimes \mathbf{1}  + \sigma_2 \otimes \sigma_3 \right)
\end{equation}

\begin{equation}
\Lambda^{24}_a = 
\left(
\begin{array}{cccc}
 0 &  0 & 0 & 0  \\
 0 &  0 & 0 & -i  \\
0  & 0  & 0 & 0 \\
0 & i & 0 & 0  
\end{array}
\right) = \frac{1}{2} \left( \sigma_2 \otimes \mathbf{1}  - \sigma_2 \otimes \sigma_3 \right)
\end{equation}

\begin{equation}
\Lambda^{14}_a = 
\left(
\begin{array}{cccc}
 0 &  0 & 0 & -i  \\
 0 &  0 & 0 & 0  \\
0  & 0  & 0 & 0 \\
i & 0 & 0 & 0  
\end{array}
\right) = \frac{1}{2} \left( \sigma_1 \otimes \sigma_2 + \sigma_2 \otimes \sigma_1 \right)
\end{equation}

\begin{equation}
\Lambda^{23}_a = 
\left(
\begin{array}{cccc}
 0 &  0 & 0 & 0  \\
 0 &  0 & -i & 0  \\
0  & i  & 0 & 0 \\
0 & 0 & 0 & 0  
\end{array}
\right) = \frac{1}{2} \left( \sigma_2 \otimes \sigma_1 -  \sigma_1 \otimes \sigma_2 \right)
\end{equation}

\begin{equation}
\Lambda^1 = 
\left(
\begin{array}{cccc}
 1 &  0 & 0 & 0  \\
 0 &  -1 & 0 & 0  \\
0  & 0  & 0 & 0 \\
0 & 0 & 0 & 0  
\end{array}
\right) = \frac{1}{2} \left( \mathbf{1} \otimes \sigma_3 + \sigma_3 \otimes \sigma_3 \right)
\end{equation}

\begin{equation}
\Lambda^2 = \frac{1}{\sqrt{3}}
\left(
\begin{array}{cccc}
 1 &  0 & 0 & 0  \\
 0 &  1 & 0 & 0  \\
0  & 0  & -2 & 0 \\
0 & 0 & 0 & 0  
\end{array}
\right)  = \frac{1}{\sqrt{3}} \left[ \sigma_3 \otimes \mathbf{1}  + \frac{1}{2} \left( \sigma_3 \otimes \sigma_3 - \mathbf{1}  \otimes \sigma_3  \right) \right]
\end{equation}

\begin{equation}
\Lambda^3 =  \frac{1}{\sqrt{6}}
\left(
\begin{array}{cccc}
 1 &  0 & 0 & 0  \\
 0 &  1 & 0 & 0  \\
0  & 0  & 1 & 0 \\
0 & 0 & 0 & -3  
\end{array}
\right)  = \frac{1}{\sqrt{6}} \left( \sigma_3 \otimes \mathbf{1}  - \sigma_3 \otimes \sigma_3 + \mathbf{1}  \otimes \sigma_3 \right)
\end{equation}

\bibliography{entropy}

\end{document}